\documentclass[a4paper]{spie}  

\usepackage[]{graphicx}
\usepackage[]{amsmath}
\usepackage{mathtools}
\usepackage{amssymb}
\usepackage[colorlinks=true,linkcolor=blue,citecolor=blue]{hyperref}

\title{Exoplanet direct imaging in ground-based conditions on THD2 bench.} 

\author{A. Potier\supit{a}, P. Baudoz\supit{a}, R. Galicher\supit{a}, E. Huby\supit{a}, G. Singh\supit{a}
\skiplinehalf
\supit{a} LESIA, Observatoire de Paris, Universit\'e PSL, CNRS, Sorbonne Universit\'e, Universit\'e de Paris, 5 place Jules Janssen, 92195 Meudon, France
}

\authorinfo{Further author information: Send correspondence to axel.potier@obspm.fr or pierre.baudoz@obspm.fr}

\begin{document} 
  \maketitle 

\begin{abstract}
The next generation of ground-based instruments aims to break through the knowledge we have on exoplanets by imaging circumstellar environments always closer to the stars. However, direct imaging requires an AO system and high-contrast techniques like a coronagraph to reject the diffracted light of an observed star and an additional wavefront sensor to control quasi-static aberrations, including the non common path aberrations. To observe faint objects, a focal plane wavefront sensor with a sub-nanometric wavefront control capability is required.\\
\noindent
In the past few years, we developed the THD2 bench which is a testbed for high-contrast imaging techniques, working in visible and near infrared wavelengths and currently reaching contrast levels lower than 1e-8 under space-like simulated conditions. We recently added a turbulence wheel on the optical path which simulates the residuals given by a typical extreme adaptive optics system and we tested several ways to remove quasi-statics speckles.\\
\noindent
One way to estimate the aberrations is a method called pair-wise probing where we record few images with known-shapes we apply on the adaptive optics deformable mirror. Once estimated, we seek to minimize the focal-plane electric field by an algorithm called Electric Field Conjugation.\\
\noindent
In this paper, we present the first results obtained on the THD2 bench using these two techniques together in turbulent conditions. We then compare the achieved performance with the one expected when all the quasi-static speckles are corrected. 
\end{abstract}



\section{INTRODUCTION}
Direct imaging of stellar environments is a powerful technique to discover new objects like circumstellar disks or for long-orbit exoplanets detection and characterization. However, the proximity between those objects and their stars added to the difference of luminosity require specific solutions. First, a telescope with a large primary mirror improves the detection probability of an exoplanet closer to the star. Currently, the difficulty to launch large mirrors to space makes essential the use of ground-based instrument. Secondly, a coronagraph is mandatory to reject the light of an observed star and to image fainter objects in its closest environment. Nevertheless, coronagraphs offer the best performance in the absence of any phase or amplitude aberrations. That is why the ground-based instruments dedicated to high-contrast imaging such as SPHERE/VLT (Beuzit et al.~\cite{Beuzit2019}) and GPI/Gemini (Macintosh et al.~\cite{Macintosh2015}) are equipped with extreme AO facilities. Yet, the Non Common Path Aberrations (NCPAs) which are not corrected by the AO system are limiting the wavefront correction to residuals of about 50nm rms.
These aberrations could be first calibrated on the internal source of the instrument. However, since the NCPAs are quasi-static and evolve slowly during the night, a regular calibration during on-sky observations would provide better contrast performance. If a few methods were tested so far by calibrating the NCPAs on the instrument internal source as the SCC (Galicher et al~\cite{Galicher2019}), the EFC (Matthews et al.\cite{Matthews2017}) or the speckle nulling (Bottom et al.~\cite{Bottom2016}), only the speckle nulling technique demonstrated an on-sky correction with SCEXAO/Subaru (Martinache et al.~\cite{Martinache2014}). As the former method shows moderate results, we decided to study the performance of the Electric Field Conjugation (EFC) in the context of turbulences.

\noindent
Sect.~\ref{sec:WSC} reminds the principle of the pair wise probing technique (PW) as a focal plane wavefront sensor associated with an EFC controller. We then describe in Sect.~\ref{sec:ExperimentalSetup} how it was implemented on the THD2 testbed simulating ground-based conditions.

\section{Wavefront sensing and control}
\label{sec:WSC}
In this section, we present the focal plane wavefront sensor and the wavefront controller we use in Sect.~\ref{sec:ExperimentalSetup} to control the quasi-static aberrations.

\subsection{Fourier numerical model of a high-contrast imaging instrument}
\subsubsection{Global model}
\label{subsubsec:GlobalModel}
Controlling the NCPAs first requires a focal plane wavefront sensor which either retrieves phase and amplitude aberrations in the pupil plane like ZELDA (Vigan et al.~\cite{Vigan2018}) and COFFEE (Herscovici et al.~\cite{Herscovici2018}) or retrieves the real and imaginary part of the electric field in the focal plane like the Self Coherent Camera (Mazoyer et al.~\cite{Mazoyer2013}) or PW. As the controller described in Sect.~\ref{subsec:EFC} and used in Sect.~\ref{sec:ExperimentalSetup} aims to generate a dark hole by minimizing the focal plane electric field, we decided to implement the PW strategy.
This method was first developped in Give'On et al.~\cite{GiveOn2007SPIE} . It defines $E_f$ the electric field in the detector plane as a function of the aberrations upstream the coronagraph:
\begin{equation}
\label{eq:totalelectricfield}
E_f=C\left[Ae^{\alpha+i(\beta+\phi)}\right],
\end{equation}
where $\alpha$ and $\beta$ are the upstream amplitude and phase aberration and $\phi$ is the phase wavefront deformation induced by a deformable mirror (DM) located in the instrument pupil plane. $C$ is the linear operator which represents the electric field propagation in a typical high-contrast instrument equiped with a phase mask coronagraph M and a Lyot stop situated in a following pupil plane
\begin{equation}
\label{eq:propmodel}
\begin{aligned}
C(E)= &\mathcal{F}\left[\mathcal{F}^{-1} \left[\text{M}\times \mathcal{F}(E)\right]\times \text{L}\right]\\
=& \left[\text{M}\times\mathcal{F}(E)\right]*\mathcal{F}(L),
\end{aligned}
\end{equation}
where $\mathcal{F}$ and $\mathcal{F}^{-1}$ are the Fourier transform and the inverse Fourier transform respectively.
When we use a perfect coronagraph with no upstream aberrations, the detector electric field vanishes. Yet, in the presence of aberrations, part of the stellar light goes through the Lyot stop to create bright speckles on the detector.
If these aberrations are NCPA, we can assume to be in a low aberration regime. We also suppose small deformations of the DM with respect to the wavelength in the rest of this paper. Therefore, $E_f$ in Eq.~\ref{eq:totalelectricfield} can be separated into two parts
\begin{equation}
\label{eq:totalelectricfield2}
\begin{aligned}
E_f&=C\left[Ae^{\alpha+i\beta}\right]+iC\left[A\phi\right]\\
&= E_S+E_{DM},
\end{aligned}
\end{equation}
where $E_S$ and $E_{DM}$ are the detector electric field induced by the upstream aberrations and the DM respectively.

\subsubsection{Model under turbulence}
\label{subsubsec:TurbulentModel}
In the former section, we did not make any assumptions on the dynamics of the upstream wavefront. However, aberrations on a ground-based instrument are a combination of static and quasi-static aberrations and aberrations produced by the atmospheric turbulence which are evolving quickly in time. In the case of an instrument equipped with extreme adaptive optics facilities, these strong fast varying aberrations are flattened by the AO system. However, a small part of turbulent aberrations called AO residuals remains. In that case, we can split $\beta$ in Eq.~\ref{eq:totalelectricfield} into two different terms to write the upstream pupil electric field $E_{pup}$ as
\begin{equation}
\label{eq:champpup}
E_{pup}=Ae^{\alpha+i(\beta_0+\beta_1(t)+\phi)},
\end{equation}
where $\beta_0$ is the static phase aberration (constant in time) whereas $\beta_1$ represents the AO residuals evolving quickly during an observation. Assuming a certain time of integration, the field $E_{pup}$ averages over N independant phase screens and $E_{pup}$ becomes
\begin{equation}
\label{eq:moychamppup}
<E_{pup}>_N=\text{E}(E_{pup})+\text{V}(<E_{pup}>_N),
\end{equation}
where $\text{E}(E_{pup})$ represents the mathematical expectation of $E_{pup}$. For an infinite exposure time, $\beta_0$ stays constant whereas the mathematical expectation of $e^{i\beta_1(t)}$ is $e^{-\sigma_1^2/2}$ where $\sigma_1$ is the time and spatial standard deviation of $\beta_1(t)$ assuming it is ergodic. Therefore, we can write
\begin{equation}
\label{eq:esperance}
\text{E}(E_{pup})=Ae^{\alpha+i(\beta_0+\phi)}e^{-\sigma_1^2/2}.
\end{equation}
In Eq.~\ref{eq:moychamppup}, V represents the variance. For $N$ independant upstream phase screens, we can write that
\begin{equation}
\label{eq:averagevar}
\text{V}(<E_{pup}>_N)=\frac{\text{V}(E_{pup})}{N}.
\end{equation}
According to Eq.~\ref{eq:esperance} and \ref{eq:averagevar}, $\text{V}(<E_{pup}>_N)$ becomes
\begin{equation}
\label{eq:variance}
\begin{aligned}
\text{V}(<E_{pup}>_N)&=\frac{1}{N}\left[\text{E}(|E_{pup}|^2)-\left(\text{E}(|E_{pup}|)\right)^2\right]\\
&=\frac{1}{N}\left[\text{E}\left((Ae^\alpha e^{i(\phi+\beta_0+\beta_1)})(Ae^\alpha e^{i(\phi+\beta_0+\beta_1)})^*\right)-(Ae^\alpha e^{i(\beta_0+\phi)}e^{-\sigma_1^2/2})(Ae^\alpha e^{i(\beta_0+\phi)}e^{-\sigma_1^2/2})^*\right]\\
&=\frac{1}{N}\left[Ae^\alpha-Ae^\alpha e^{-\sigma_1^2}\right]\\
&=\frac{Ae^\alpha(1-e^{\sigma_1^2})}{N}.
\end{aligned}
\end{equation}
Eventually, assuming small quasi static aberrations and small AO residuals with respect to the wavelength, we use Eq.~\ref{eq:esperance} and Eq.~\ref{eq:variance} to simplify the Eq.~\ref{eq:moychamppup} as
\begin{equation}
\label{eq:EFPupTurbu}
<E_{pup}>_N=Ae^{\alpha}\left(1+i(\beta_0+\phi)\right)+\frac{Ae^\alpha\sigma_1^2}{N}.
\end{equation} 
The former equation means that acquiring long exposure images averages out the turbulent residual phase maps. The longer the exposure, the lower is the importance of phase residuals in the detector electric field $E_f$ and Eq.~\ref{eq:totalelectricfield2} becomes
\begin{equation}
\label{eq:EFDetectorTurbu}
E_f=E_{S_0}+E_{DM}+\epsilon ,
\end{equation}
where $E_{S_0}$ is the detector electric field induced by the quasi-static aberrations whereas $\epsilon$ represents the small contribution of the AO residuals to the detector electric field when acquiring long exposure images. The term $\epsilon$ becomes negligeable if the AO contribution in Eq.~\ref{eq:EFPupTurbu} is small when compared to the contribution of the quasi-static. This is valid when the number of averaged phase screens is high enough:
\begin{equation}
N>>\frac{\sigma_1^2}{\beta_0+\phi}.
\end{equation}

\subsection{Pair-Wise probing}
\label{subsec:PairWiseProbing}
PW is a useful focal plane wavefront sensor. It is too slow to detect AO residuals but it could be used with long exposure images to measure the NCPAs (see in Sect.~\ref{subsubsec:TurbulentModel}). PW is implemented by recording pairs of images when applying known-shapes $\psi$ on the DM. Those applied patterns are the so-called probes which modulate the speckle intensity in the detector image to create an image $I_m$. According to Eq.~\ref{eq:EFDetectorTurbu} in the case of a long enough exposure, the turbulent contribution $\epsilon$ becomes negligeable in the correction zone and $I_m$ can be written as
\begin{equation}
I_m=|E_{S_0}+iC[A\psi_m]|^2.
\end{equation}
Pushing and pulling a probe in this linear regime cancels out the quadratic terms of $I_m$ to retrieve a linear combination of the real part $\Re$ and the imaginary part $\Im$ of the searched electric field $E_S$
\begin{equation}
I_m^+-I_m^-=4(\Re(E_{S_0})\Re(iC[A\psi_m])+\Im(E_{S_0})\Im(iC[A\psi_m])).
\end{equation}
In this case, a pair of probes is insufficient to separate the two terms of this complex field. However, the use of $k$ different pair of probes allows to lift this degenacy. By representing it in matrix algebra,
\begin{eqnarray}
\begin{bmatrix}
\label{eq:estimation} 
 I_1^+-I_1^- \\
. \\
. \\
. \\
I_k^+-I_k^- 
\end{bmatrix}_{(i,j)}=4
\begin{bmatrix}
\Re (iC[A\psi_1]) & \Im (iC[A\psi_1]) \\
. & . \\
. & . \\
. & . \\
\Re (iC[A\psi_k]) & \Im (iC[A\psi_k]) 
\end{bmatrix}_{(i,j)}
\begin{bmatrix}
 \Re (E_{S_0}) \\
 \Im (E_{S_0})
\end{bmatrix}_{(i,j)} ,
\end{eqnarray}
\noindent
we may retrieve the real and imaginary part of the detector plane electric field when inverting Eq.~\ref{eq:estimation}. The reader may note this inverse problem stays well-posed only if at least two probes $m$ and $n$ are modulating the speckle intensity differently on each pixel.
\begin{eqnarray}
\label{eq:EnoughDiversity}
\Re(iC[A\psi_m])\Im(iC[A\psi_n])-\Re(iC[A\psi_n])\Im(iC[A\psi_m])\neq 0 .
\end{eqnarray}
If the condition~\ref{eq:EnoughDiversity} is verified, we can write $E_S$ on each pixel as a linear combination of the probe images difference:
\begin{eqnarray}
\label{eq:estimation2}
\begin{bmatrix}
 \Re (E_{S_0}) \\
 \Im (E_{S_0})
\end{bmatrix}_{(i,j)}=\frac{1}{4}
\begin{bmatrix}
\Re (iC[A\psi_1]) & \Im (iC[A\psi_1]) \\
. & . \\
. & . \\
. & . \\
\Re (iC[A\psi_k]) & \Im (iC[A\psi_k])
\end{bmatrix}^{\dagger}_{(i,j)}
\begin{bmatrix}
 I_1^+-I_1^- \\
. \\
. \\
. \\
I_k^+-I_k^-
\end{bmatrix}_{(i,j)} .
\end{eqnarray}

\subsection{Electric Field Conjugation}
\label{subsec:EFC}
After estimating the static wavefront in the focal plane with PW, we aim to control the quasi-static aberrations $E_{S_0}$. We assume that the electric field induced by the DM in the science detector is a linear combination of the actuator movements:
\begin{equation}
\label{eq:LinEFC}
E_{DM}=G\bar{a} , 
\end{equation}
where $G$ is a linear transformation between the DM actuator voltages $\bar{a}$ and the focal plane electric field. $G$ is implemented numerically, by propagating the effect of each actuator movement on the focal plane through the $C$ operator. We then look for the best commands to apply in order to minimize the speckle intensity in a region of the image called Dark Hole (DH). We decide to use the Electric Field Conjugation method which seek to minimize the distance
\begin{equation}
d^2_{EFC}=||E_{DM}+E_{S_0}||^2 .
\end{equation}
This optimization is solved by calculating the inverse matrix of $G$ with a Singular Value Decomposition (SVD). Therefore, the voltages to apply to the DM are the result of a matrix multiplication between the command matrix and the PW result electric field
\begin{equation}
\label{eq:pasactionneurs}
\bar{a}=-g[\Re (G)^\frown\Im (G)]^{\dagger}[\Re (E_{S_0})^\frown\Im (E_{S_0})] ,
\end{equation}
where $^\frown$ means concatenate and $^\dagger$ means pseudo inverse.

\section{Experimental Setup}
\label{sec:ExperimentalSetup}
\subsection{The THD2 testbed}
\label{subsec:THD2}
\begin{figure}[htp]
   \centering
   \includegraphics[width=13cm]{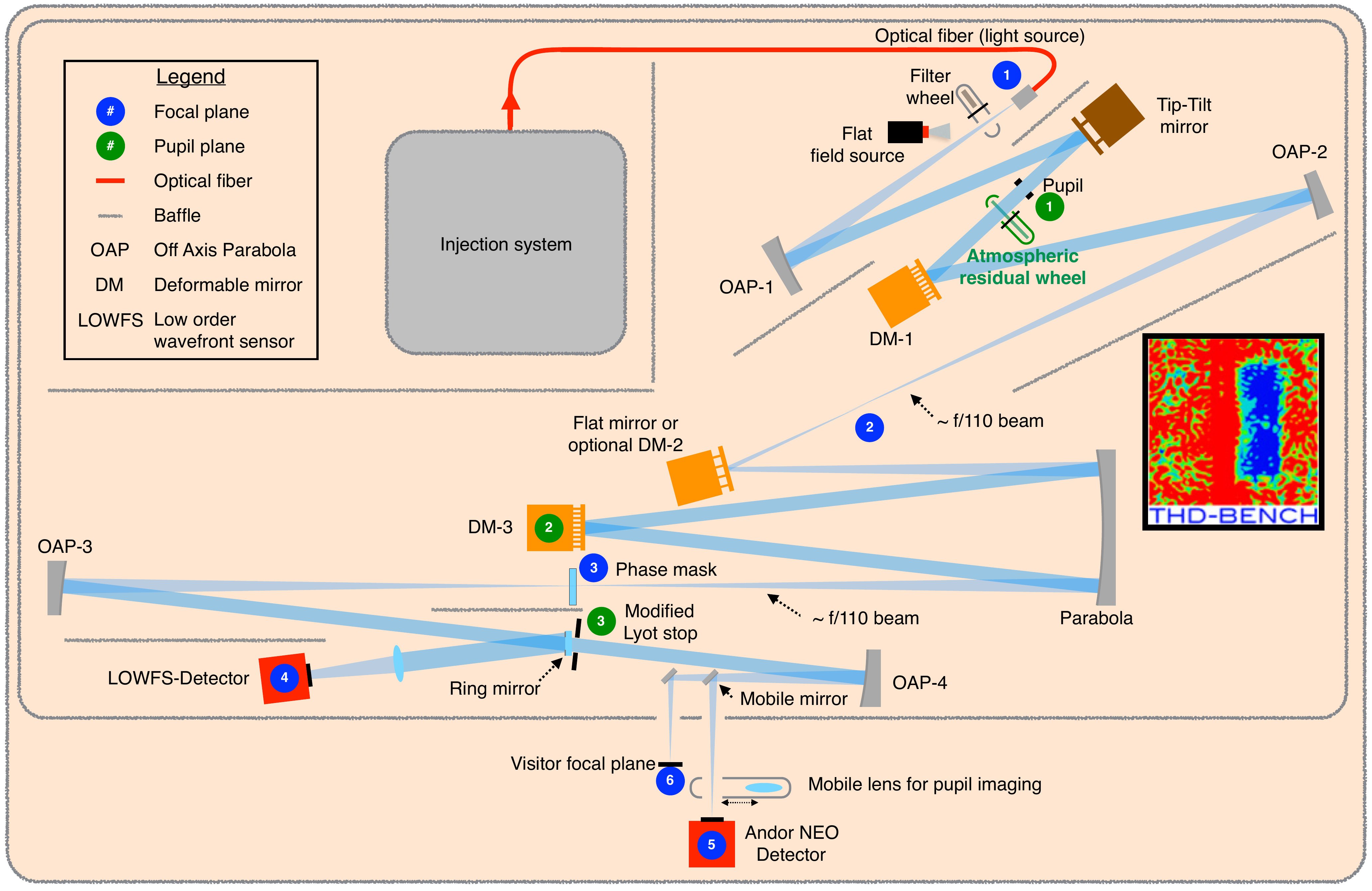}
   \caption{Optical configuration of the THD2 testbed.}
              \label{fig:THD2}%
\end{figure}

The THD2 bench is a testbed located at LESIA and dedicated to high-contrast imaging. Although it was originally developped for space-based applications, we recently added a turbulence wheel to test NCPA compensation with a ground-based instrument. The testbed is detailed in Baudoz et al.~\cite{Baudoz2018} and its optical components are represented in Fig.~\ref{fig:THD2}. For the rest of this paper, it is used with a monochromatic light source of wavelength 783.25nm and an entrance pupil of diameter 8.23mm. The coronagraph is composed of a FQPM mask diffracting the simulated stellar light out of the Lyot stop of diameter 8.00mm. A 2-electron read-out noise sCMOS camera record the coronagraph images. Two servo loops are used in the following part. First, a low-order wavefront sensor located in the focal plane 4 (LOWFS, Singh et al.~\cite{Singh2019}) collects the light from the reflective Lyot-stop to control the Tip-Tilt mirror at 100Hz. It is used to keep the star Point Spread Function (PSF) centered on the FQPM cross. Then, the quasi-static aberrations are controlled through the EFC method by the 32x32 Boston-Micromachine (DM3) settled in a pupil plane. These aberrations are directly probed with PW in the sCMOS camera.

\subsection{The turbulence wheel}
For this particular experiment, the testbed is used in a "ground configuration". In this case, a turbulent wheel is inserted in the instrument right after the entrance pupil. This rotating phase plate, described in Singh et al.~\cite{Singh2019} , simulates the post-AO residuals of the SPHERE instrument in the infrared in optimistic conditions for the current SPHERE performance. The power spectral density (PSD) of the aberrations follows a power law as $f^{-4/3}$ below the DM correction cut-off (20 resolution element) and a power law as $f^{-11/3}$ above. The standard deviation of the aberration is about 40nm rms. At the position of the entrance pupil, we move the plate at a steady velocity of 6.3mm.s$^{-1}$ which corresponds to 6.1m.s$^{-1}$ when the pupil is scaled to an 8m telescope.

\subsection{Results on THD2 testbed} 
\begin{figure}[htp]
   \centering
   \includegraphics[width=17cm]{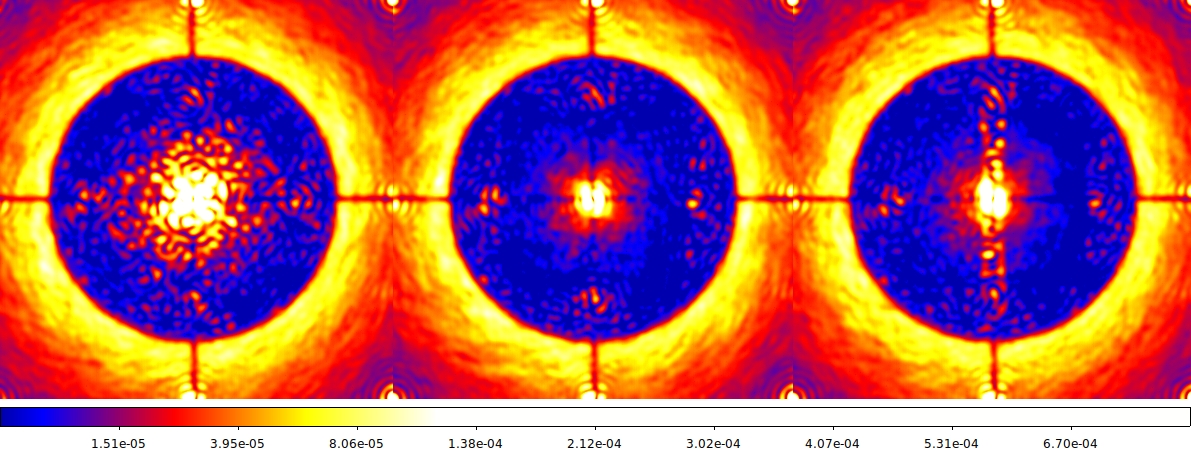}
   \caption{Coronagraphic images with 18s exposure time. Left: no quasi-static correction is applied. Center: Result of a 5 iterations correction in a 25$\lambda/D\times25\lambda/D$ FDH. Right: Result of a 3 iteration correction in a rectangular HDH of size 10$\lambda/D\times25\lambda/D$, located on the right of the star.}
              \label{fig:ResultImages}%
\end{figure}

\begin{figure}[htp]
   \begin{center}
   \includegraphics[width=12cm]{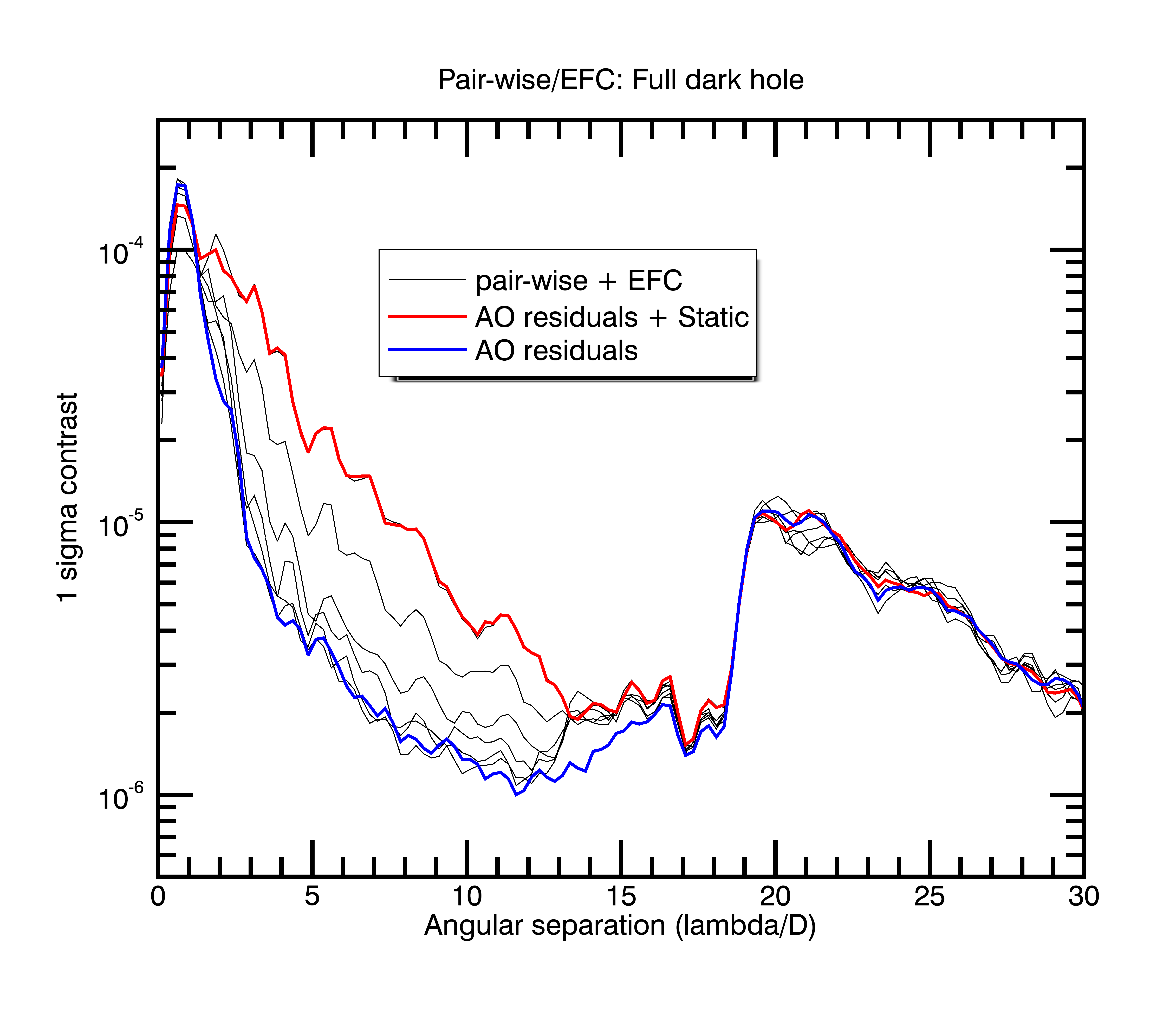}
   \includegraphics[width=12cm]{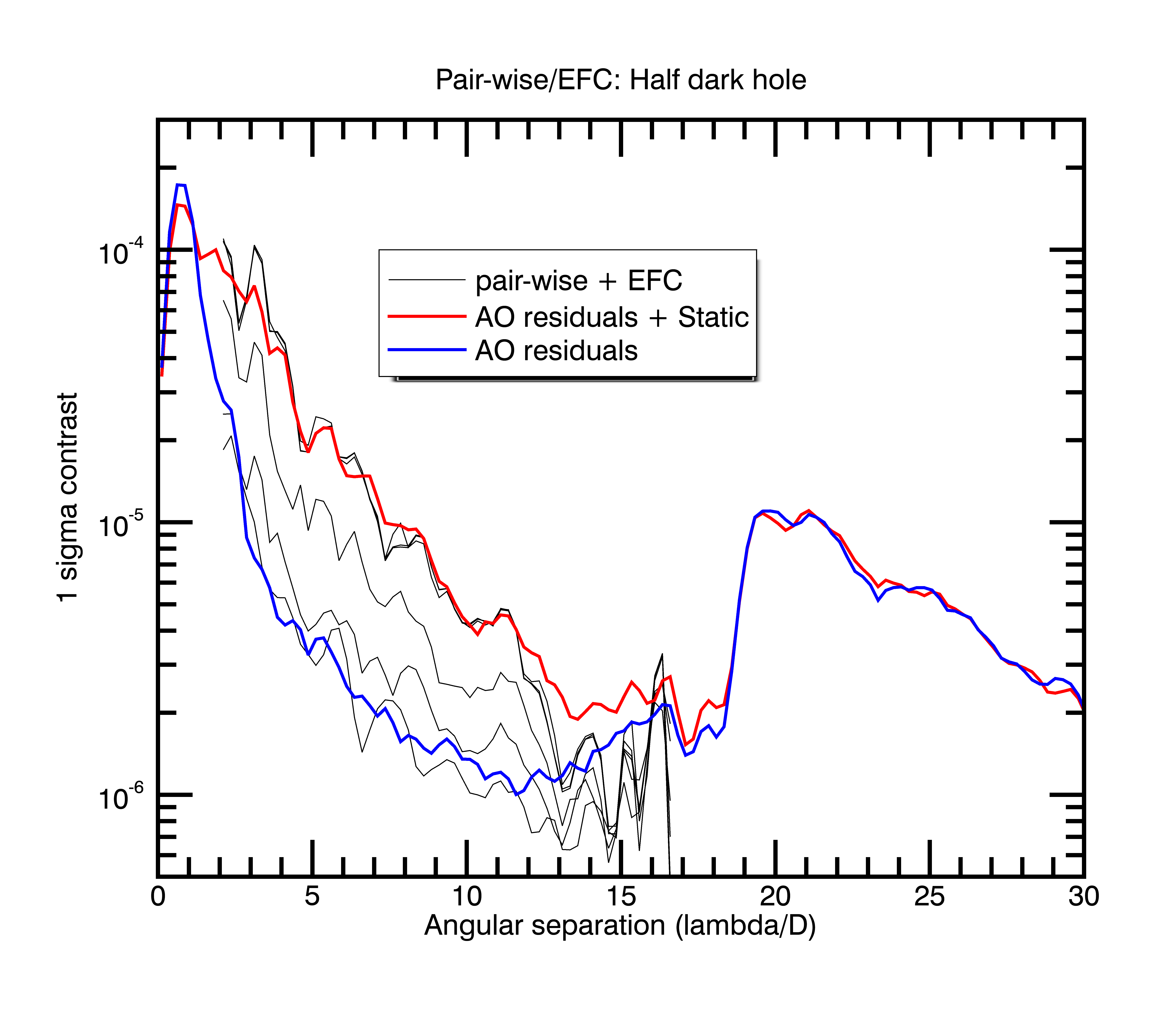}
   \end{center}
   
   \caption{1-$\sigma$ contrast performance when applying PW+EFC in a squared FDH of size 25$\lambda/D\times25\lambda/D$ (top) and in a rectangular HDH of size 10$\lambda/D\times25\lambda/D$ (bottom).}
              \label{fig:ResultContrast}%
\end{figure}

In this section, we demonstrate that the combination of PW and EFC is able to correct for quasi-static aberrations when turbulent conditions also affect the coronagraphic images. In order to test the performances of this combined technique, we decided to use the 1-sigma contrast metric where we calculate the azimutal standard deviation of the intensity in the coronagraphic image normalized by the maximum of the non-coronagraphic PSF.\\
We first remove the rotating phase plate from the optical path to calibrate the quasi-static aberrations with the usual PW+EFC close loop in a full dark hole (FDH) configuration. We then inserted the phase plate in the optical path and rotate it. Eventually, we recorded a set of long exposure images of 18s with the sCMOS camera to ensure that the turbulent wavefront averages. We plotted the 1-sigma contrast for the obtained image in blue in Fig.~\ref{fig:ResultContrast} to represent the best achievable contrast in the conditions of a perfect control for phase static aberrations (assuming the rotation of the wheel to fully average the phase which is not true since it has a finite size).\\
Then, we applied about 5nm of phase aberrations with the DM3. The PSD of these aberrations decrease as $f^{-1}$ which corresponds to the typical NCPAs remaining in SPHERE after a previous compensation on the internal source. Again, we recorded a set of 18s image exposure while the phase plate was rotating to obtain the image in Fig.~\ref{fig:ResultImages} (Left). The reader may notice the theoretical SPHERE AO cut-off on this image above 20 cycles per pupil. Inside the corrected region, bright speckles remains due to the quasi-static aberration just applied on the DM3. The corresponding contrast curve obtained is plotted in Fig.~\ref{fig:ResultContrast}, in red.\\
We then look for correcting the quasi-static speckles while rotating the turbulence wheel to simulate an on-sky NCPA correction. First, we estimate the aberrations with PW as described in Sect.~\ref{subsec:PairWiseProbing} recording 18s exposure probe images. The probes used are 3 different bumps of actuators of 30nm peak to valley. Once estimated, we seek to control the static aberrations in a squared FDH of size 25$\lambda/D\times25\lambda/D$ with EFC using only DM3. After 5 iterations of PW+EFC, we recorded the image at the center in Fig.~\ref{fig:ResultImages} where the reader notice the disappearance of the bright speckles in the dark hole region. Note that DM3 owns less actuators in the pupil than the SPHERE DM which induces a narrower accessible region of correction in the THD2 testbed. It leaves uncorrected speckles between the DM cut-off and the phase plate cut-off. In a science instrument, the AO DM and the DM used for NCPA correction are the same and this effect would disappear. The contrast for such a correction is plotted in grey on the top image in Fig.~\ref{fig:ResultContrast} for each iteration. It shows that the generated DH reaches the floor of a perfect correction (blue curve) meaning that the phase aberration are corrected after 5 iterations. The correction of quasi-static aberrations is therefore performed in about 9 minutes in these conditions.\\
Eventually, we tested a half dark hole (HDH) correction to estimate PW+EFC ability to correct both phase and amplitude aberrations during observations. We started from the same static aberrations as for the FDH correction. We used the same probes for the focal plane electric field estimation with the same exposure time. However, the EFC correction is done in a HDH going from -12$\lambda/D$ to 12$\lambda/D$ in one direction and from 2$\lambda/D$ to 12$\lambda/D$ in the other direction. We represent the image obtained after 3 iterations on the right in Fig.~\ref{fig:ResultImages} and we plot each iteration contrast obtained in the HDH region in grey in the bottom plot of Fig.~\ref{fig:ResultContrast}. This plot shows better performance in a HDH correction than for the FDH correction. Indeed, both phase and amplitude aberrations can be corrected in a HDH with only one DM located in a pupil plane whereas only phase aberrations are corrected in a FDH. This means that the speckles generated from amplitude aberrations remain in the FDH correction and limit the instrument performance.

\section{Conclusion}
This paper presents a strategy to correct for NCPAs in a high-contrast imaging instrument on ground such as SPHERE or GPI. We described the PW+EFC method we used under a phase screen simulating residual turbulence. If the AO system is good enough to give residual aberrations lower than the wavelength, it shows that such NCPA correction techniques - first developped in space conditions - can also be used from ground by averaging the turbulence in long exposure images. We then presented laboratory results of such a test of NCPA correction. It demonstrates that we are able to actively correct for both phase and amplitude NCPAs during on-sky observations applying phase diversities. In the conditions we tested on the testbed (3 to 5 iterations and 18s per images), it requires only 10 minutes to generate a DH cleaned of speckles induced by the NCPAs.
This work is promising for the future of ground-based high-contrast imaging instruments. Intruments in design SPHERE+ and GPI2 could easily use this active NCPA correction process whithout modifying their hardware.

\acknowledgments
The authors would like to thank the Centre National d’Etudes Spatiales (CNES), DIM-ACAV+ and the Marie Skłodowska-Curie research fellowship programme for their financial support. Founded in 1961, the CNES is the french government agency responsible for shaping and implementing France space policy in Europe. DIM-ACAV+ is a Ile-de-France region funding and supporting research around Paris in the fields of astrophysics and the conditions for life appearance. Marie Skłodowska-Curie actions support researchers at all stages of their careers. It funds Garima Singh research under grant No.  798909. This work was also supported by the Action Spécifique Haute Résolution Angulaire (ASHRA) of CNRS/INSU co-funded by CNES.

\bibliography{bib_GS}   
\bibliographystyle{spiebib}   

\end{document}